\documentclass{kluwer}    % Specifies the document style.
\usepackage[dvips]{epsfig}
\newdisplay{guess}{Conjecture}
\newcommand{\msun}{$M_{\odot}$}
\newcommand{\etal}{{\it et al. }}
\newcommand{\beq}{\begin{equation}}
\newcommand{\eeq}{\end{equation}}

\begin{document}                                                                                   
\begin{article}
\begin{opening}         
\title{Trapped Disk Oscillations and Stable QPOs in Micro-\\quasars} 
\author{Xue-Bing \surname{Wu}$^{1,2}$}  
\runningauthor{Xue-Bing Wu}
\runningtitle{Trapped Disk Oscillations and Stable QPOs in Microquasars}
\institute{$^1$Max-Planck-Institut f\"ur Astrophysik, 
       Karl-Schwarzschild Str.~1, D-85741 Garching, Germany\\
       $^2$Beijing Astronomical Observatory, Chinese Academy of
       Sciences, Beijing 100012;
       CAS-PKU Joint Beijing Astrophysical Center, Beijing 100871, China 
}
%\date{September 15, 2000}

\begin{abstract}
The observed quasi-periodic oscillations (QPOs) in low-mass X-ray
binaries have been suggested to be probably related to the 
characteristic %frequencies of 
oscillation modes in the inner part of
a relativistic accretion disk. In this paper we study the 
trapped feature of these oscillation modes by considering the 
radial perturbations and viscosity effects. 
We find that the trapped oscillations are amplified significantly when the viscosity parameter
increases and the trapped size varies when the radial 
wavenumber and Mach number change. Our results support that the stable
 hundred-hertz QPOs and their low fractional rms amplitudes found in two Galactic 
microquasars may be explained by some trapped 
non-damping oscillation modes in the inner part of accretion
disks around  rotating black holes.

\end{abstract}
\keywords{accretion, accretion disks, black hole physics, X-ray: stars}
\end{opening}           

\section{Introduction}  
                    % Produces section heading.  Lower-level
                    % sections are begun with similar 
                    % \subsection and \subsubsection commands.

In recent years, hundred-hertz QPOs %quasi-periodic oscillations
have been observed  in several black hole X-ray binaries with {\it
RXTE} satellite. These high-frequency QPOs are probably similar to
the kilohertz QPOs  observed frequently in a dozen  neutron
star low-mass X-ray binaries. 
However,  the high
frequency QPOs in two
Galactic microquasars,  GRO J1655-40 and GRS 1915+105,  are very stable in frequency at 300 Hz and
67 Hz respectively (Morgan, Remillard \&
Greiner 1997; Remillard \etal 1999). These stable QPOs have also very
low fractional rms amplitude ($\sim 1 \%$). Theoretical models involving
frame-dragging (Cui, Zhang \& Chen 1998), diskoseismic g-mode
(Nowak \etal 1997), and periastro precession (Stella, Vietri \& Morsink 1999)
have been proposed to explain these stable QPOs.  But the origin of
them is still uncertain.

Recently Psaltis \& Norman (2000; hearafter PN) suggested that the observed QPOs in
%neutron star and black hole
 X-ray binaries may be related to the strong resonances peaked at some
characteristic
oscillation frequencies in the inner part of accretion disks. 
  However, 
they
introduced a simplified term for angular momentum transfer
%in their equations 
and  did not %were unable to
discuss  %the effects of viscosity on 
the damping and growth rates of
oscillation modes. %in the disk. 
It is also important to check
%if these oscillation modes are damping modes and 
if some of these modes
can be trapped  in the inner  disk
%.  It is also interesting to investigate 
and  if the low
fractional rms amplitude of QPOs observed in two microquasars can be explained by the
trapped feature of oscillation modes. 
%These are the aims of the current paper.

\section{Non-damping Oscillation Modes in Accretion Disk}

%We consider a geometrically thin, non-self-gravitating accretion disk
%around a rotating central compact object. 
The basic equations of a geometrically thin accretion disk are the
same as those in PN except that the viscosity
term in the angular momentum equation is now expressed as:
$N_\phi=(1/\rho r)\partial (\nu\rho r^3 d\Omega/dr)/\partial r$.
%, where $\rho$ is the density and $\Omega$ is the angular velocity. 
The viscosity is assumed to be described by the usual $\alpha$ prescription,
namely, $\nu=\alpha c_{\rm s}
H=\alpha c_{\rm s}^2/\Omega$.
%, where $c_{\rm s}^2$ is the local sound
%speed and $H$ is the disk height. 
%For simplicity we have ignored the
%radial and vertical viscous force.
Like in PN, we consider the general relativistic effects in an approach to
specify the the gravitational potential $\psi$  through
the dependences  of angular velocity ($\Omega$), radial and vertical
epicyclic frequencies ($\kappa$ and $\Omega_\perp$) on the mass, radius
and the spin parameter of the central object.
We consider small perturbations to all physical quantities in the form
of $Q=Q_0+\delta Q
e^{i(\omega t - k r - m\phi)}$, where $k$ and $m$ are radial and
azimuthal wavenumbers. We also assume that the 
perturbations do not alter the local sound speed and the disk is
isothermal in the vertical direction. Under the local
approximation $k r>>1$, we can obtain four perturbed equations
%according to four basic equations.
%By taking the determinant of the coefficients in the perturbed
%equations to zero, we can obtain 
and derive the dispersion relation as
$$
{\bar{\sigma}}^4+\alpha\epsilon^2{\bar{\sigma}}^3 +({\bar{\kappa}}^2+\epsilon^2
+n{\bar{\Omega}}_\perp^2){\bar{\sigma}}^2 - \epsilon [m\frac{H}{r}
(\frac{{\bar{\kappa}}^2}{2}-2)+2\alpha \epsilon
(\frac{{\bar{\kappa}}^2}{2}-2)
$$
\beq
~~~~~~~~~-\alpha \epsilon^3 -n\epsilon {\bar{\Omega}}_\perp^2]{\bar{\sigma}}
+n{\bar{\kappa}}^2 {\bar{\Omega}}_\perp^2 =0.
\eeq
Here we have defined %some dimensionless quantities given by 
$\bar{\sigma}=i(\omega -ku_r-m \Omega)/\Omega$, $\bar{\kappa}=\kappa/\Omega$, 
 $\bar{\Omega}_\perp=\Omega_\perp/\Omega$, and $\epsilon = k H$. The
value of $n$ represents the number of nodes of the oscillations in the
vertical direction. 
%The above dispersion relation is very helpful to determine the
%stability properties and oscillation features in accretion disks.
%In the limits of low $k$ and $\alpha=0$, this
%relation is identical to that obtained by Kato (1989, 1990) and is similar. 
%to that derived in the model of 
%``Diskoseismology'' (see Nowak \& Wagoner 1992). 

\begin{figure} % fig 1
%\vskip -0.3cm
\hspace{-0.5cm}
\psfig{file=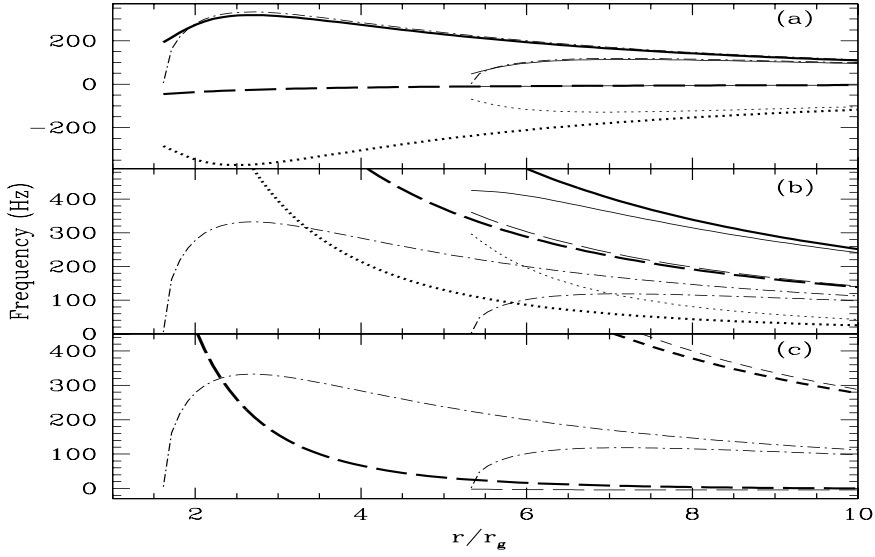,width=12.5cm, height=8cm}
\vskip -0.4cm
\caption[]{Oscillation frequencies of non-damping
modes when $kH=0.1$ and $|u_r|/c_s=0.3$. The weighted and light lines
represent the cases with black hole spin parameter
$a_*=0.98$ and 0.2. The panel (a), (b) and (c) correspond
to (m, n)=(0, 0), (1, 0) and (1,1). In (a) and (b) the solid, dotted
and dashed lines  represent
two acoustic modes and the neutral mode. In (c) the long and short dashed
lines represent two corrugation modes. The dot-dashed lines show the
radial epicyclic frequencies. The black hole mass was taken to be
7\msun and $r_g$ was defined as $GM/c^2$.}
%\label{penG}
\end{figure}

\begin{figure} % fig 1
%\vskip -0.3cm
\hspace{-0.5cm}
\psfig{file=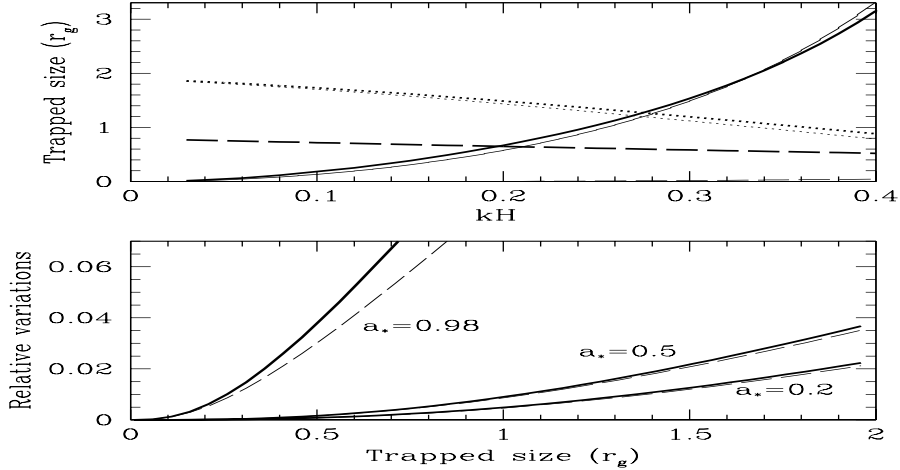,width=12.5cm, height=6.5cm}
\caption[]{Upper panel: Trapped sizes of three
oscillation modes. The solid, dotted and
dashed lines represent the lower-frequency acoustic modes in the cases of (m, n)
=(0,0) and (1,0), and the lower-frequency corrugation mode in the case
of (m, n)=(1,1). The weighted and light lines correspond to the cases
with $a_*=0.98$ and 0.2. Lower panel: Relative variations caused by the trapped oscillations. The solid and dashed lines correspond to
the cases with accretion rate of $10^{17} g/s$ and $10^{18} g/s$.
}
%\label{penG}
\end{figure}

%The  characteristic frequencies
%and the damping (or growth) rates of oscillation modes can be
%obtained 
By solving the dispersion relation (1),
 we found that the characteristic frequencies of oscillation modes 
depend on the
radial wavenumber and radius significantly but 
are nearly independent of the 
viscosity. However, the growth (or damping) rates of these modes
depend strongly on  the viscosity. According to their
non-damping feature, we found that the most possible
oscillation modes in the disk are the acoustic modes with approximate frequencies of
$ ku_r\pm\sqrt{\kappa^2+k^2c_{\rm s}^2}$ (in the case of $m=0$ and
$n=0$) and  $ ku_r+\Omega\pm\sqrt{\kappa^2+k^2c_{\rm s}^2}$ (in the
case of $m=1$ and $n=0$), neutral modes with frequency of $ku_r$ (in
the case of $m=0$ and $n=0$) and $
ku_r+\Omega$ (in the case of $m=1$ and $n=0$), and the corrugation
modes with approximate
frequencies of $ku_r+\Omega \pm \Omega_\perp$ (in the case of $m=1$ and
$n=1$). 
In Figure 1 we show the dependences of oscillation frequencies of all non-damping 
modes on the disk radius and black hole spin parameters.
% for the cases with (m, n)=(0, 0), (1, 0) and (1,
%1), respectively.
Note that the lower-frequency acoustic modes and the lower-frequency corrugation mode are
trapped in the inner part of accretion disks since $\omega^2>\kappa^2$.
The trapped sizes of these  modes depend significantly on the radial
wavenumber, Mach number
and black hole spin parameter. The relative luminosity variations caused by
the oscillations in the trapped area may be estimated using the 
expression of local flux emitted from a relativistic accretion disk.
%(Page \& Thorne 1974). 
These relative variations may be comparable to the observed fractional
rms magnitudes if the X-ray luminosity variations are
caused by the disk oscillations in the trapped area. In Figure 2 we show the trapped sizes
of the oscillation modes and the relative  luminosity variations.

\section{Conclusion}

{\it RXTE} observations on GRO J1655-40 have shown 300 Hz
QPOs with $\sim 1\%$ rms magnitude. If this stable QPO originates from the
trapped disk oscillations, from Figures 1 and 2 we see that
it may be explained by either the lower-frequency acoustic
mode in the case of $m=1$ and $n=0$ or the lower-frequency
corrugation mode in the case of $m=n=1$ and $a_*\sim 0.9$. 
However, the rapidly spinning black hole is more favored to explain the
high QPO frequency. The unambiguous determinations of the inner disk radius
and the black hole spin are still needed to confirm this explanation. Further studies
on the origin of  low-frequency QPOs and their correlation with
high-frequency QPOs, as well as the correlation between  spectral and
temporal properties, will  help us to understand more clearly the origin  of stable
QPOs in
microquasars.

%\acknowledgements
%I thank the  support from the CAS-MPG exchanging program.

% The endnotes section will be placed here.

%\theendnotes

\end{article}

\begin{thebibliography}{}

\bibitem[1997]{czc98}
Cui, W., Zhang, S.N., Chen, W.
\newblock Evidence for frame-dragging around spinning black holes in
\newblock X-ray binaries,
\newblock  {\em Ap. J.} 492:L53-57, 1998
\bibitem[1997]{morgan}
Morgan, E.H., Remillard, R.A., Greiner, J.
\newblock RXTE observations of QPOs in the black hole candidate GRS 1915+105,
\newblock {\em  Ap. J.} 482:993-1010, 1997
\bibitem[1997]{nowaket97}
Nowak, M.A., Wagoner, R.V., Begelman, M.C., Lehr, D.E. 
\newblock The 67 Hz feature in the black hole candidate GRS 1915+105 as a
\newblock possible ''diskoseismic'' mode,
\newblock {\em Ap. J.}  477:L91-94, 1997
\bibitem[2000]{psaltis00}
Psaltis, D., Norman, C.
\newblock On the origin of quasi-periodic oscillations and broad-band
\newblock noise in accreting neutron stars and black holes,
\newblock {\em Ap. J.} in press (astro-ph/0001391)
\bibitem[1999a]{remillard}
Remillard, R.A., Morgan, E.H., McClintock, J.E., Bailyn, C.D., Orosz, J.A. 
\newblock RXTE observations of 0.1-300 Hz quasi-periodic oscillations
\newblock in the microquasar QGO J1655-40,
\newblock {\em  Ap. J.} 522:397-412, 1999
\bibitem[1999a]{stella99a}
Stella, L., Vietri, M., Morskink, S.M.
\newblock Correlations in the quasi-periodic oscillation frequencies
\newblock of low-mass X-ray binaries and the relativistic precession model,
\newblock {\em Ap.J.} 524:L63-66, 1999



\end{thebibliography}
\end{document}